\documentclass[pre,twocolumn,showpacs,preprintnumbers,amsmath,amssymb]{revtex4}

\usepackage{graphicx}

\begin{document}

\preprint{APS/123-QED}

\title{A reaction-diffusion model for the growth of avascular tumor}

\author{S. C. Ferreira Junior$^1$}
 \email{silviojr@fisica.ufmg.br}

\author{M. L. Martins$^2$}
 \email{mmartins@mail.ufv.br}

\author{M. J. Vilela$^3$}

\affiliation{$^1$Departamento de F\'{\i}sica, Instituto de Ci\^encias Exatas, Universidade Federal de
Minas Gerais, CP 702, 30161-970, Belo Horizonte, MG, Brazil}

\affiliation{$^2$Departamento de F\'{\i}sica, Universidade Federal de Vi\c{c}osa, 36571-000, Vi\c{c}osa, 
MG, Brazil}

\affiliation{$^3$Departamento de Biologia Animal, Universidade Federal de Vi\c{c}osa, 36571-000, Vi\c{c}osa, 
MG, Brazil}

\date{\today}

\begin{abstract}
A nutrient-limited model for avascular cancer growth including cell proliferation, motility 
and death is presented. The model qualitatively reproduces commonly observed morphologies for primary tumors,
and the simulated patterns are characterized by its gyration radius, total number of cancer cells, and number
of cells on tumor periphery. These very distinct morphological patterns follow Gompertz growth curves, but exhibit
different scaling laws for their surfaces. Also, the simulated tumors incorporate a spatial structure composed of a
central necrotic core, an inner rim of quiescent cells and a narrow outer shell of proliferating cells in agreement
with biological data. Finally, our results indicate that the competition for nutrients among normal and cancer cells
may be a determinant factor in generating papillary tumor morphology.
\end{abstract}


\maketitle

\section{Introduction}
Cancer is a disease derived, with few exceptions, from mutations on single
somatic cells that disregard the normal controls of proliferation, invade
adjacent normal tissues and give rise to secondary tumors (metastasis) on sites
different from its primary origin~\cite{Hanahan}. Although cancers are
extremely diverse and heterogeneous, a small number of pivotal steps associated
to both deregulated cell proliferation and suppressed cell death is required
for the development of any and all tumors. Indeed, all neoplasms evolve
accordingly to an universal scheme of progression~\cite{Clark,Evan}. Neoplastic
cells accumulate a series of genetic or epigenetic changes along the tumor
progression in response to natural selection and as an integrated defense
program against stress situations similar to the response of bacterial colony
facing severe and sustained threats~\cite{Lucien}. But, unlike of the bacterial
growth, tumor progression involves a complex network of interactions among
cancer cells and its host microenvironment~\cite{Liotta}. It is well known that
stressed bacterial colonies can develop diffusion-limited fractal
patterns~\cite{Golding,Kozlovsky}. Recently, normal and tumor cell patterns
{\it in vivo} and {\it in vitro} were characterised by their fractal dimensions
and cluster size distribution functions~\cite{Cross, Sato, Mendes}, reinforcing
the great current interest in the search for basic principles of growth in
living organisms, which are the most complex and challenging self-organized
systems. In particular, for cancer growth, one of the most aggressive phenomena
in biology, numerous mathematical models have been recently investigated.
Examples include studies, based on classical reaction-diffusion equations, of
the growth of tumor spheroids \cite{Pettet}, cancer  evolution and its
interation with the immune system \cite{Bellomo}, and the fundamental problem
of tumor angiogenesis  \cite{Levine,DeAngelis}.

Recently, we~\cite{Jr1,Jr2} proposed a diffusion-limited model to  simulate the
growth of carcinoma ``in situ'' in which cell proliferation, motility and death
are locally regulated by the concentration of growth factors produced by each
cancer cell. This model was able to generate compact, connected and
disconnected morphologies which progress in time according to Gompertz growth
curves, and for which the tumor gyration radii scale as in the Eden model for
an asymptotically large number of cells. These features were shown to be
independent of the detailed functional form of the microscopic growth rules. In
contrast, the structure of the tumor border is influenced by the nature of the
growth rules as indicated by the different scaling laws for the number of
peripheral cancer cells. In particular, for disconnected and connected patterns
the surface widths scale with exponents similar to those observed in bacterial
colonies. Although the similarities between the simulated and histological
tumor patterns were encouraging, the model was unable to generate papillary and
ramified morphologies found in many of epithelial cancers and trichoblastomas.
In consequence, we were lead to investigate the role of nutrient competition in
neoplasic development, a biological feature excluded in our previous model but
central in the growth of fractal bacterial colonies. Indeed, cancer cells
subvert the evolutionary adaptations to multicellularity and revert to a
largely nutrient-limited style of growth.

In this paper we analyze the avascular cancer growth in a model including cell
proliferation, motility and death as well as competition for nutrients among
normal and cancer cells. The cell actions (division, migration and death) are
locally controlled by the nutrient concentration field. In Section 2 we
introduce the cancer growth model. In Section 3 the simulational results for
the tumor patterns, growth curves and scaling laws are discussed. In Section 4
the diffusion of  growth factors secreted by the cancer cells are added to the
nutrient-limited model. Finally, we draw some conclusions in Section 5.

\section{The cancer growth model}
The basic biological principles included in the model are cell proliferation,
motility and death and competition for nutrients among normal and cancer cells.
Nutrients (oxigen, amino acids, glucose, metal ions, etc.) diffuse from a
capillary vessel through the tissue towards the individuals (normal and cancer)
cells. Under restricted nutrient supply the growth rate of cancer cells is
limited by its ability to compete for nutrients with the normal cells. In our
model the division, migration and death of each cancer cell is controlled by
the nutrient concentration in its local microenvironment.

\subsection{The tissue}
The studied system consists of a tissue fed by a single capillary vessel. The
tissue is represented by a square lattice of size $(L+1)\times (L+1)$ and
lattice  constant $\Delta$. The capillary vessel, localized at the top of the
lattice at $x=0$, is the unique source from which nutrients diffuse through the
tissue towards the individuals cells. Although a tumor mass is composed of
different cell subpopulations~\cite{Clark}, we shall consider only three types:
normal, cancer and tumor necrotic cells. Any site, with  coordinates
$\vec{x}=(i\Delta,j\Delta)$, $i,j=0,1,2,\ldots,L$, is occupied by only one of
these cell types. In contrast to normal cells, one or more cancer cells can
pile up in a given site. In turn, tumor necrotic cells are inert and, for
simplicity, will be considered always as a single dead cell. Thus, each lattice
site can be thought of as a group of actual cells in which the normal, necrotic
and cancer cell populations assume one of the possible values
$\sigma_n(\vec{x},t)= \sigma_d(\vec{x},t)=0,1$ and
$\sigma_c(\vec{x},t)=0,1,2,\ldots$, respectively. As initial ``seed'' a single
cancer cell in the half of the lattice ($x=L\Delta/2$) and at a distance $Y$
from the capillary vessel is introduced in the normal tissue, in agreement with
the theory of the clonal origin of cancer~\cite{Nowell}. Periodic boundary
conditions along the horizontal axis are used. The row $i=0$ represents a
capillary vessel and the sites with $i=L+1$ constitute the external border of
the tissue.

\subsection{The nutrients}
As considered by Scalerandi {\it et al.}~\cite{Scalerandi1}, we assume that
dividing cancer cells are  especially vulnerable to some critical nutrients
such as iron, essential for DNA synthesis and, therefore, for cell division.
The many other nutrients necessary for eucariotic cells are supposed to affect
mainly the motility and death of the cancer cells. So, the nutrients are
divided into two groups: essential and non-essential for cell proliferation,
described by the concentration fields $N(\vec{x},t)$ and $M(\vec{x},t)$,
respectively. However, it is assumed that both nutrient types have the same
diffusion coefficients and consumption rates by the normal cells. These
concentration fields obey the diffusion equations:

\begin{eqnarray}
\frac{\partial{N(\vec{x},t)}}{\partial{t}}=D\nabla^2 N(\vec{x},t)- \gamma
N(\vec{x},t) \sigma_n (\vec{x},t)\nonumber\\ -\lambda_N \gamma N(\vec{x},t)
\sigma_c (\vec{x},t) \end{eqnarray}
and

\begin{eqnarray}
\frac{\partial{M(\vec{x},t)}}{\partial{t}}=D\nabla^2 M(\vec{x},t)-\gamma
M(\vec{x},t) \sigma_n (\vec{x},t)\nonumber\\ -\lambda_M \gamma M(\vec{x},t)
\sigma_c (\vec{x},t) \end{eqnarray}
in which the nutrient absorption terms are proportional to the cell populations
present in each site, and it is assumed differentiated nutrient consumption
rates for normal and cancer cells by factors $\lambda_N$ and $\lambda_M$. It is
important to notice that the present model assumes the simplest form for the
nutrient diffusion phenomena, i. e., linear equations with constant
coefficients. Also, $\lambda_N > \lambda_M$ is used, reflecting the larger
cancer cells affinity for essential nutrients.

The boundary conditions satisfied by the nutrient concentration fields are
$N(x=0)=M(x=0)=K_0$, representing the continuous and fixed supply of nutrients
provided by the capillary vessel; $N(y=0)=N(y=L\Delta)$ and
$M(y=0)=M(y=L\Delta)$, corresponding to the periodic boundary conditions along
the x-axis; at last, Neumann boundary conditions,
$\partial{N(x=L\Delta)}/\partial{y}=\partial{M(x=L\Delta)}/\partial{y}=0$, are
imposed to the border of the tissue. The hypothesis that a blood vessel
provides a fixed nutrient supply to the cells in a tissue is a simplification,
which neglects the complex response of the vascular system to metabolic changes
of cell behavior~\cite{Scalerandi2}.

In order to reduce the  number of parameters in equations (1) and (2), the new
dimensionless variables are defined:

\begin{eqnarray}
t^\prime=\frac{Dt}{\Delta^2}\;, ~ \vec{x^\prime}=\frac{\vec{x}}{\Delta}\;, ~ N^\prime=\frac{N}{K_0}\;,
\nonumber\\ M^\prime=\frac{M}{K_0}\;, ~ \alpha=\Delta \sqrt{\frac{\gamma}{D}}\;.
\end{eqnarray}
Using these new variables in Eqs. (1) and (2) and omitting the primes we obtain

\begin{equation}
\frac{\partial{N}}{\partial{t}}=\nabla^2 N-\alpha^2 N \sigma_n -\lambda_N \alpha^2 N \sigma_c 
\end{equation}
and

\begin{equation}
\frac{\partial{M}}{\partial{t}}=\nabla^2 M-\alpha^2 M \sigma_n -\lambda_M \alpha^2 M \sigma_c
\end{equation}
for the diffusion equations  describing the nutrients concentration fields. In
addition, the boundary condition on the capillary vessel becomes
$N(x=0)=M(x=0)=1$ and a value $\Delta=1$ is defined.

\subsection{Cell dynamics}
Each tumor cell can be selected at random, with equal probability, and carry out one of three actions: 

(1) {\it division}. Cancer cells divide by mitosis with  probability $P_{div}$.
If the chosen cell is inside the tumor, its daughter will pile up at that site,
and $\sigma_c(\vec{x}) \rightarrow \sigma_c(\vec{x})+1$. Otherwise, if the
selected cell is on the tumor border, its daughter cell will occupy at random
one of their nearest neighbor sites $\vec{x^\prime}$ containing a normal or a
necrotic cell and, therefore, $\sigma_c(\vec{x^\prime})=1$ and
$\sigma_{n,d}(\vec{x^\prime})=0$. The mitotic probability $P_{div}$ is
determined by the concentration per cancer cell of the essential nutrients $N$
present on the microenvironment of the selected cell:

\begin{equation}
P_{div}(\vec{x})=1-\exp\left[- \left( \frac{N}{\sigma_c \; \theta_{div}} \right)^2 \right].
\end{equation}
The Gaussian term is included in order to produce a sigmoid curve saturated to
the unity, and the model parameter $\theta_{div}$ controls the shape of this
sigmoid.

(2) {\it migration}. Cancer cells migrate with probability $P_{mov}$.  A
selected cell inside the tumor, at a site $\vec{x_i}$, will move to a nearest
neighbor site $\vec{x ^\prime}$ chosen at random. Thus, $\sigma_c(\vec{x
^\prime}) \rightarrow \sigma_c(\vec{x ^\prime})+1$ and, clearly,
$\sigma_c(\vec{x}) \rightarrow \sigma_c(\vec{x})-1$. Otherwise, if the selected
cell is on the tumor border, the invasion of a normal or necrotic nearest
neighbor site will be dependent on the number of cancer cells present in the
selected site. If in this site there is a single cancer cell, it migrates by
interchanging its position with that of the invaded one. If there are other
cancer cells in the same site of the one selected to move, the migrating cell
will occupy the position of the normal or necrotic nearest neighbor cell,
which, in turn, disappears. In terms of cell populations the migration of a
cell on the tumor border corresponds to the following operations:
$\sigma_c(\vec{x ^\prime})=1$, $\sigma_c(\vec{x}) \rightarrow
\sigma_c(\vec{x})-1$, $\sigma_{n,d}(\vec{x ^\prime})=0$ and
$\sigma_{n,d}(\vec{x})=1$ if $\sigma_c(\vec{x})=1$. The probability of cell
migration $P_{mov}$ has the same functional form of $P_{div}$, but depends on
the concentration of the non-essential nutrients $M$ present on the
microenvironment of the selected cell and increases with the local population
of cancer cells. So,

\begin{equation}
P_{mov}(\vec{x})=1-\exp\left[- \sigma_c \;\left( \frac{M}{\theta_{mov}} \right)^2 \right]
\end{equation}
with the model parameter $\theta_{mov}$ controlling the shape of this sigmoid.

(3) {\it cell death}. Cancer cells die transforming in a  necrotic cell with
probability $P_{del}$. Thus, $\sigma_c(\vec{x}) \rightarrow
\sigma_c(\vec{x})-1$ and $\sigma_d(\vec{x})=1$ when $\sigma_c$ vanishes. The
cell death probability $P_{del}$ is determined by the concentration per cancer
cells of the non-essential nutrients $M$ present on the microenvironment of the
selected cell:

\begin{equation}
P_{del}(\vec{x})=\exp\left[-\left( \frac{M}{\sigma_c \; \theta_{del}} \right)^2 \right],
\end{equation}
a Gaussian distribution whose variance depends on the model parameter $\theta_{del}$.

The cell dynamics rules used in our model take into account  that, as the
cancer growth progress, cell migration increases near the border of the tumor
due to the high availability of nutrients and the increase in the number of
cancer cells, which release a series of enzymes (collagenases,
metalloproteinases, etc.) responsible by the progressive destruction of the
extracellular matrix. Also, in the regions where there is a high population
density and an ineffective supply of nutrients via diffusion processes, the
cell division is inhibited and, at the same time, the probability of cell death
increases. But, under these rules cell growth and migration is possible even
inside the tumor. Finally, the model parameters $\theta_{div}$, $\theta_{mov}$
and $\theta_{del}$, which characterize the cancer cells response to nutrient
concentrations and embody complex genetic and metabolic processes, should be
interpreted in terms of the underlying biochemistry and molecular biology, an
still open problem. The other three model parameters $\alpha$, $\lambda_N$ and
$\lambda_M$, associated to the consumption of essential and non-essential
nutrients for cell proliferation by the normal and cancer cells, should be more
easily determined from biological experiments.

It is worthwhile to notice that from the point of view of the so-called kinetic
cellular theory, which provides a general framework for the statistical
description of the population dynamics of interacting cells \cite{Bellomo}, the
local probabilities $P_{div}$, $P_{mov}$ and $P_{del}$ can be thought as an
effective kinetic cellular model.

\subsection{Computer implementation}

The growth model simulations were implemented using the following procedure.
At each time step, the diffusion Eqs. (4) and (5) are numerically solved in the
stationary state through relaxation methods, providing the nutrient
concentration at any lattice site. Then, $N_{C}(t)$ cancer cells are
sequantially selected at random with equal probability. For each one of them, a
tentative action (division, death or movement) is chosen at random with equal
probability and the time is incremented by $\Delta t=1/N_{C}(t)$. The selected
cell action will be implemented or not according to the correspondent local
probabilities determined by Eqs. (6), (7) or (8). If the selected cell divides
or die, therefore changing the number of cancer cells which consume nutrients,
we solve the diffusion equations in a small neighborhood of linear size $l=20$
centered in the altered site. This is done in order to take into account these
local perturbations and to speed up the computer algorithm, since the number of
numerical iterations need to solve the diffusion equation is proportional to
$L^2$. At the end of this sequence of $N_C(t)$ tentatives, a new time step
begins and the entire procedure (solution of the diffusion equations and
application of the cell dynamics) is iterated. The simulations stop if any
tumor cell reaches the capillary vessel. In all the simulations, the exact
solutions for the stationary diffusion equatios in the absence of tumor cells
were used for the nutrient concentration fields at $t=0$.

\section{Simulational results}
In Figure 1 the most commonly observed morphologies in tumor growth such as
papillary,  compact and disconnected are shown. Disconnected patterns, typical
of round cells neoplasies such as lymphoma, mastocytoma, and plasmacytoma,
correspond to transient simulated patterns in which cancer cells have high
motility but a low mitotic rate. In turn, if the cell migration is very small
and the high mitotic rate demands large amounts of essential nutrients, then
the simulated patterns exhibit finger-like shapes similar to the papillary
morphologies found in epithelial tumors, such as basal and skin cells
carcinomas, and hepatomas. If nutrient availability in the tissue is further
reduced by an increasing cell consumption, the papillary-like patterns become
progressively thin and similar to the chords or filaments of cells which
constitute one of the hallmarks of the trichoblastoma morphology. Finally,
under high nutrient supply, which means low cell consumption of both nutrient
types, the simulated patterns are compact such as those observed for solid
tumors.

Typical patterns generated by the present model are shown in Figure 2. As one
can see, nonspherical morphologies growing towards the capillary vessel are
observed, according to the rigorous results for moving boundary problems of
cancer growth \cite{Friedman}. The nutrient consumption by normal and cancer
cells, controlled by the model parameters $\alpha$, $\lambda_N$ and
$\lambda_M$, plays a central role in morphology determination. For small values
of these parameters, corresponding to growth conditions in which individual
cells demand small nutrients supplies, the patterns tend to be compact and
circular. However, if the mitotic rate of cancer cells is small due to the
large amount of nutrients demanded for cell division, generating a significant
nutrient competition, these compact patterns progressively assume
papillary-like morphologies. At high nutrient consumption rates these papillary
patterns become the rule and, for low cancer cell division, continuously
transform in thin tips, filaments or chords of cells. Also, the smaller the
$\alpha$ and $\lambda_N$ values are, the larger is the fraction of necrotic
cancer cells for a fixed cellular response leading to cell death (controlled by
$\theta_{del}$). In particular, this fraction is smaller for papillary patterns
than compact ones, suggesting that the optimal growth morphology under strong
nutrients limitation is fractal. Finally, the higher cell migration is, more
homogeneous are the patterns, faster is the tumor growth, and smaller is the
fraction of necrotic cells.

The tumor patterns generated by the present model were characterized by its
gyration radius $R_g$,  total number of cancer cells $N_C$, and number of cells
on tumor periphery $S$ (including the surface of holes, if any). The gyration
radius $R_g$ is defined as

\begin{equation}
R_g=\left(  \frac{1}{n} \sum_{i=1}^{n} r_{i}^2 \right)^2
\end{equation}
where $n$ is the number of sites occupied by the pattern (cancer or necrotic
cells) and $r_i$ is the  distance of the occupied site $i$ from the tumor mass
center. These quantities could be related to clinically important criteria such
as progress curves, rate of growth (volumetric doubling time) at given radii,
proliferative and necrotic fractions of the tumor. In medicine, these data are
used to determine tumor's malignancy and its prognosis. The obtained results
are summarized in Table 1.

Despite of the detailed microscopic mechanisms of growth, for all the simulated
patterns the  progress in time of cancer cell populations follows a Gompertz
curve

\begin{equation}
N_C(t)=A \exp \left[ - \exp(-k(t-t_c))\right]
\end{equation}
as one can see in Figure 3. Moreover, the tumor gyration radius $R_g$ as well
as the number  of peripheral cells $S$ also exhibit a Gompertz growth, contrary
to the linear regimes observed for $R_g$ and even $S$ in our previous non
nutrient-limited model~\cite{Jr2}. So, the present model reveals that the
Gompertz law of growth for the cancer cell populations and tumor size is a
robust emergent feature of cancer dynamics under nutrient competition. It is
important to notice that linear or power law fittings for the growth in time of
$R_g$ and $S$ can also be very satisfactory for several simulated patterns,
therefore explaining our previous findings~\cite{Jr2} and the recent
experimental observations of brain tumors grown {\it in vitro}~\cite{Bru}.
However, only the Gompertz law can provide good fittings for all the three
quantities $N_C$, $R_g$ and $S$.

On the other hand, as a function of the total number of cancer cells both $R_g$
and $S$ obey  power law scaling given by $R_g \sim N_C^{\nu}$ and $S \sim
N_C^{\sigma}$, respectively. For solid patterns these exponents are $\nu \sim
0.5$ and $\sigma \sim 0.5$, corresponding to effective circular and non fractal
patterns. As the nutrient consumption increases the patterns tend to
papillary-like shapes for which the exponent $\sigma$ increases towards the
value $1$ and the exponent $\nu$ varies in the range $[0.50,0.60]$, indicating
a fractal morphology for the tumor. It is important to mention that increasing
cell motility contributes to round and homogenize these patterns, progressively
destroying their fractal features. Also, the tumor growth is faster for higher
cancer cell migration.

An interesting result, shown in Figure 4, is the existence of a necrotic core
in the center  of the simulated tumors for high nutrient consumption or cell
division rates. As observed in real tumors and {\it in vitro} multicell
spheroids~\cite{Pettet}, a simulated pattern consists of three distinct
regions: a central necrotic core, an inner rim of quiescent cancer cells and a
narrow outer shell of proliferating cells. These different regions are evident
in Figure 5 where the cancer cell density and average cell division rate are
plotted for a longitudinal cut across the growth pattern. As one can see, both
the cancer cell density and mitotic rates have neat sharp maxima at the tumor
borders in front and opposed to the capillary vessel. Notice that the peaks for
division rates are significantly more narrow than those for cancer cells
density, demonstrating that the proliferative fraction of the tumor comprises
just a small part of the cancer cells localized at the tumor border.

However, in the present nutrient-limited model the disconnected patterns common
in cancers  of round cells (Fig. 1e) correspond to transient behaviors of the
model for low mitotic rates and high cell migration of cancer cells. Also, the
ramified morphology of a trichoblastoma seem in Fig. 6d could not be
qualitatively reproduced by the model. A worthwhile feature of this pattern is
the presence of ``leaves'', the growth of each one clearly influencing the
others. The simple introduction of cell motility sensitive to the nutrients
gradients failed to generate stationary disconnected or ramified patterns.
Therefore, in addition to the nutrient field, it appears that a chemotactic
interaction among cancer cells guiding their migration must be in action. The
nature of this biological interaction and the results of its simulation will be
focused in the next section.

\section{Growth factors}
From the biological point of view the chemotactic response to growth factors
released by cancer cells seems  to be, in addition to nutrient supply, another
central feature in cancer development. The reciprocal influence among cancer
cells mediated by autocrine and paracrine growth factors, motility factors,
etc. influence the microenvironment of each cancer cell and hence its division
and migration. Thus, both diffusive fields (nutrients and growth factors)
determine the local probabilities for cancer cells divide, migrate and die.

\subsection{The model}
In order to investigate the role of growth factors in a nutrient-limited
growth, we simplify  our model by considering a single nutrient field described
by the diffusion equation:

\begin{equation}
\frac{\partial{N}}{\partial{t}}=D_N\nabla^2 N-\gamma N \sigma_n -\lambda \gamma N \sigma_c ~,
\end{equation}
in which $\gamma$ and $\lambda \gamma$ are the nutrients consumption  rates of
normal and cancer cells, respectively. The boundary conditions are the same
described in Section 2. In turn, the growth factors (GFs) concentration obeys
the diffusion equation:

\begin{equation}
\frac{\partial{G}}{\partial{t}}=D_G\nabla^2 G - k^2 G + \Gamma \sigma_c N (G_M-G) ~,
\end{equation}
which includes the natural degradation of GFs, also imposing a characteristic
length $\sim 1/k$ for GFs diffusion, and a production term increasing linearly
with the local nutrient concentration up to a saturation value $G_M$.
Therefore, we are assuming that the release of GFs involves complex metabolic
processes supported by nutrient consumption. The boundary conditions satisfied
by the GFs concentration field is $G(\vec{x},t)=0$ at a large distance
($d>2/k$) of the tumour border .

Again, the number of parameters in equations (9) and (10) can be reduced by
using  the new dimensionless variables

\begin{eqnarray}
t^\prime=\frac{D_N t}{\Delta^2}\;, ~ \vec{x^\prime}=\frac{\vec{x}}{\Delta}\;,
~ N^\prime=\frac{N}{K_0}\;, ~ G^\prime=\frac{G}{G_M}\;, \nonumber\\ ~ \alpha=
\sqrt{\frac{\gamma \Delta^2 }{D_N}}\;,~ k^\prime=k
\sqrt{\frac{\Delta^2}{D_N}}\;, \Gamma^\prime=\frac{\Gamma \Delta^2}{D_N} \;,
D=\frac{D_G}{D_N}.
\end{eqnarray}
Using these new variables in Eqs. (9) and
(10) and omitting the primes we obtain

\begin{equation}
\frac{\partial{N}}{\partial{t}}=\nabla^2 N-\alpha^2 N \sigma_n -\lambda \alpha^2 N \sigma_c 
\end{equation}
and

\begin{equation}
\frac{\partial{G}}{\partial{t}}=D\nabla^2 G-k^2 G +\Gamma \sigma_c N (1-G)
\end{equation}
for the diffusion equations. The boundary condition for the nutrients on the
capillary vessel becomes $N(x=0)=1$ and a value $\Delta=1$ is defined. In
addition, at the stationary state the parameter $D$ in Eq. (13) can be put
equal to the unity by rescaling the parameters $k$ and $\Gamma$. So, the
diffusion equations for the nutrients and GFs involve four parameters.

On the other hand, the cell dynamics has essentially the same rules used in
Section 2, but with different cell action probabilities. The single change
introduced is that after cell division the daugther cell stands at the same
site occupied by its mother. Since nutrients are essential to the large protein
and DNA synthesis necessary to cell mitosis and GFs act are mitotic inductors,
the proposed form to $P_{div}$ is:

\begin{equation}
P_{div}(\vec{x})=1-\exp \left[- \left( \frac{N}{\sigma_c} -N^* \right) \frac{G^2}{\theta_{div}^2} \right] ~.
\end{equation}
The parameter $N^*$ determines the nutrient-poor level below which  the cancer
cells reproduction is inhibited.

Cell migration involves large citoskeleton reorganizations which  consume
energy and is facilitated by GFs which destroy the extracellular matrix and the
adhesivity structures between normal cells. Thus, $P_{mov}$ is written as:

\begin{equation}
P_{mov}(\vec{x},\vec{x^\prime})=1-\exp \left[ \frac{N(\vec{x})G(\vec{x})
[G(\vec{x})-G(\vec{x^\prime})]}{\theta_{mov}} \right] ~,
\end{equation}
implying that a cell migrates in a gradient-sensitive way towards sites  where
the GFs concentration is lower than that in its starting point.

Finally, cell death is produced by the lack of nutrients:

\begin{equation}
P_{del}(\vec{x})=\exp\left[- \left( \frac{N}{\sigma_c \; \theta_{del}} \right)^2 \right].
\end{equation}

\subsection{Results}
In Figure 6 typical compact, ramified and disconnected simulated patterns are
shown. The ramified structure shown in Fig. 6b should be compared with the
pattern of a trichoblastoma exhibited in Fig. 6d. In contrast to the compact
fingers of the papillary patterns of the previous section, in these ramified
morphologies the tumor has fjords and tips similar to those abserved in DLA
patterns. We emphasize that without chemotactic signaling among cancer cells
the nutrient-limited model can not generate stationary disconnected patterns.
The reason is that in average GFs driven cell migration outwards, promoting the
tumor expansion and, in consequence, generating disconnected patterns for high
cell motility. Again, the patterns were characterized by its gyration radius
$R_g$, total number of cancer cells $N_C$, and number of cells on tumor
periphery $S$. Essentially the same results were obtained for cancer progress
in time, scaling relations and spatial structures exhibiting a central necrotic
core, an inner rim of quiescent cells and a narrow outer shell of proliferating
cells.

\section{Conclusions}

A nutrient-limited model for the growth of avascular tumors was investigated by
numerical  simulations. In its original version, cell proliferation, motility
and death were locally regulated by the concentration of nutrients supplied by
a distant capillary vessel. These nutrients were divided into two groups, the
first one associated to the usual metabolic cell needs, and the second
essential to the synthesis of proteins and nucleic acids involved in cell
division. The nutrients concentration fields were determined by solving the
diffusion equation on the square lattice modeling the primary tissue. Our
simulation results show that the progress in time of the total number of cancer
cells, tumor gyration radius and number of cells on the tumor border is
described by Gompertz curves. The generated compact and papillary or
finger-like morphologies obey different scaling laws for the number of
peripheral cancer cells. For compact patterns $S \sim N_C^{1/2}$ as in the Eden
model, whereas for papillary patterns the exponent in the power-law increases
towards unity as the nutrient consumption increases, indicating a fractal
morphology for the tumor. Since in this model version the cell migration is not
driven by chemotactic signals secreted by the cancer cells, cell motility
contributes to round and homogenize the growth patterns. Also, the simulated
tumors incorporate a spatial structure composed of a central necrotic core, an
inner rim of quiescent cells and a narrow outer shell of proliferating cells in
agreement with biological data.

In order to simulate disconnected and ramified tumor patterns, typical of round
cells  tumors and trichoblastoma, a chemotactic interaction among cancer cells
mediated by growth factors was added to the competition for nutrients. Again,
similar results were obtained for cancer progress in time and scaling
relations. Thus, the Gompertz law emerges as a robust feature of the
nutrient-limited model of cancer growth.

Despite the encouraging results for the progress curves, scaling laws and
growth  patterns morphologies, including nonspherical symmetries, the major
feature of the present work is an attempt in connecting the macroscopic
diffusion equations for nutrients and/or growth factors to cell response and
interactions at the microscopic scaling through an effective kinetic cellular
model. Indeed, the local probabilities $P_{div}$, $P_{mov}$ and $P_{del}$
describe in a stochastic way the dynamical processes occuring in cell
populations as a response to the nutrient and growth factors diffusive fields.
Finally, further studies on angiogenesis, therapy and tumor-host interactions
using variants of the present model are under progress.

\begin{acknowledgements}
The authors would like to thank Dr. Lissandro Concei\c{c}\~ao from the UFV
Veterinary  Department for kindly providing us with the histological sections
of the tumors. We are indebted to a referee for calling our attention to recent
references and for many usefull coments to improve this manuscript. This work
was partially supported by the CNPq and FAPEMIG Brazilian agencies.
\end{acknowledgements}

\clearpage
\begin{table}
\caption{Morphology, progress curves and scaling laws for the patterns generated by the nutrient-limited model}

\begin{tabular}{|c|c|c|c|c|c|c|} \hline \hline
Morphology & Characteristic & \multicolumn{3}{c|}{Growth in time} & \multicolumn{2}{c|}{Exponents} \\ \cline{3-7}
           & features       & $N_C$ & $R_g$ & $S$ & $\nu$ & $\sigma$ \\ \hline
           & low nutrient             &          &          &          &      &      \\
Compact    & consumption;             & Gompertz & Gompertz & Gompertz & 0.50 & 0.50 \\
           & low cell motility        &          &          &          &      &      \\ \hline  
	     & high nutrient            &          &          &          &  &  \\ 
Papillary  & consumption;             & Gompertz & Gompertz & Gompertz & 0.50-0.60 & 0.60-1 \\    
           & low cell motility        &          &          &          &  &  \\ \hline  
             & low mitotic rate;      &          &          &          &      &  \\  
Disconnected & high cell motility;    & Gompertz & Gompertz & Gompertz & 0.50 & 1 \\ 
             & transient behavior     &          &          &          &      &  \\  \hline  \hline
\end{tabular}
\end{table}

\clearpage

\begin{figure*}
\begin{center}
\resizebox{10cm}{15.3cm}{\includegraphics{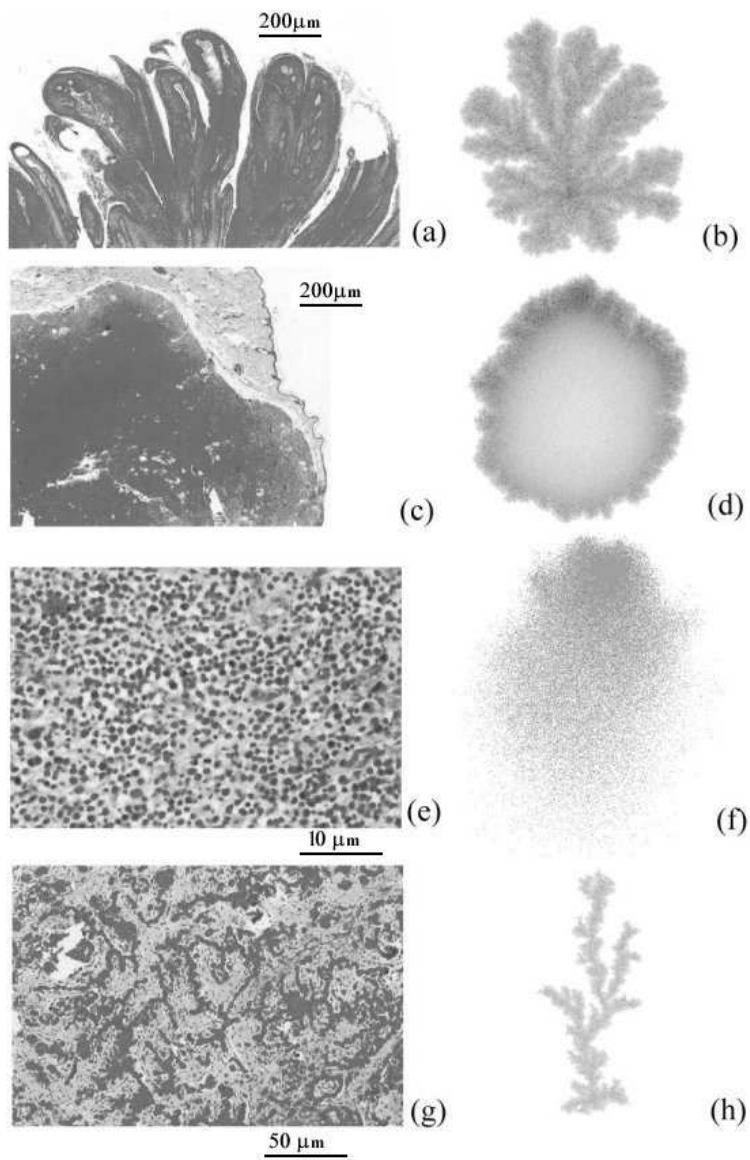}}
\end{center}
\caption{\label{fig:fig1} Commom morphologies observed in cancer growth. (a)
Papillary pattern of a scamous papyloma, (c) a compact solid basocellular
carcinoma, (e) a disconnected pattern of a plasmacytoma, and (g) characteristic
cell filaments of a trichoblastoma. All these histological patterns were
obtained from dogs. The corresponding simulated patterns are shown in (b), (d),
(f) and (h), respectively.}
\end{figure*}

\begin{figure*}
\begin{center}
\resizebox{12.7cm}{14.3cm}{\includegraphics{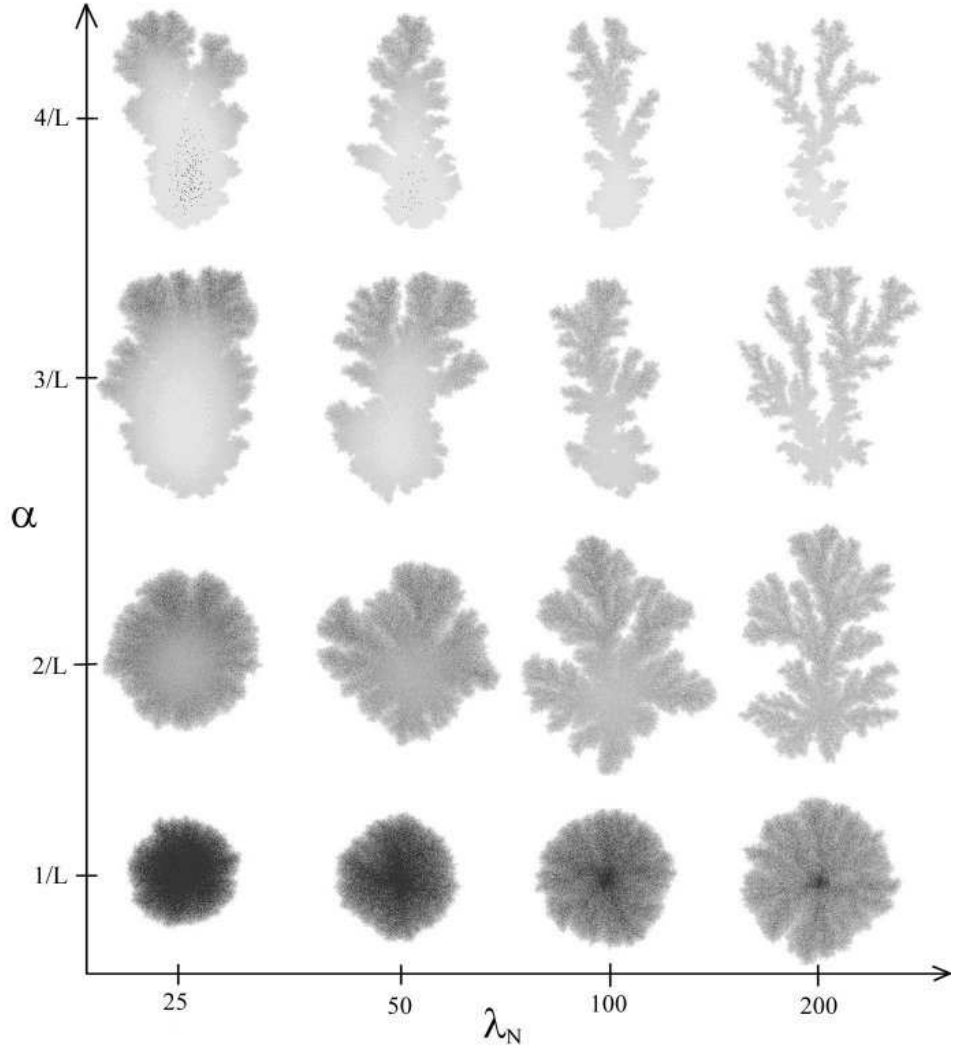}}
\end{center}
\caption{\label{fig:fig2} Simulational results of the nutrient-limited cancer
growth model. The patterns are organized as function of nutrient consumption
rate $\alpha$ for normal cells and the multiplicative factor $\lambda_N$ to the
consumption rate of mitotic essential nutrients by the cancer cells. The
remaining four parameters of the model were fixed in $\lambda_M=10$,
$\theta_{div}=0.3$, $\theta_{mov}=\infty$ (absence of cell migration) and
$\theta_{del}=0.01$. The patterns are drawn in a gray scale where the darker
regions represent higher cancer cell populations. The tissue size is $500
\times 500$, with the initial ``cancer seed'' distant $300$ sites from the
capillary. The total number of cancer cells depends on tumor morphology and
attains up to $2 \times 10^5$ for compact patterns. The simulated patterns are
compact for low $\lambda_N$ values and become papillary or finger-like for high
$\lambda_N$. For the same $\lambda_N$ the patterns are more papillary for
higher $\alpha$. Since the capillary vessel provides a fixed nutrients supply,
the dimensionless consumption rate of the normal tissue $\alpha$ set up the
levels of available resources for which cancer cells compete. So, high $\alpha$
and/or $\lambda_N$ values correspond to the limit of strong nutrient
competition.}
\end{figure*}

\begin{figure*}
\begin{center}
\resizebox{7.8cm}{7.2cm}{\includegraphics{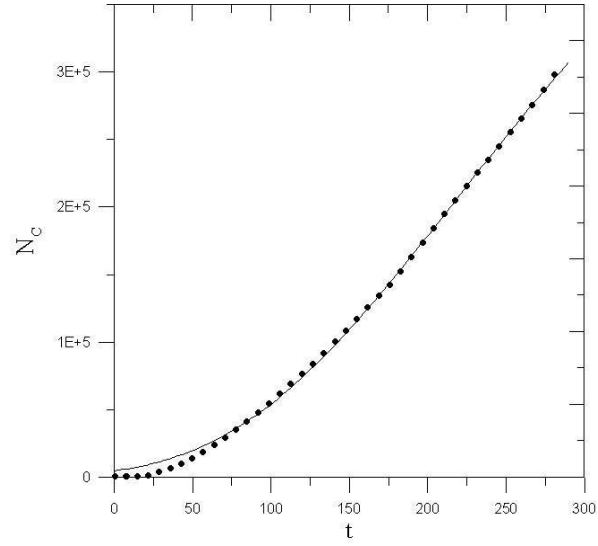}}
\end{center}
\caption{\label{fig:fig3}  Gompertz growth of cancer cell population. The
model parameters were fixed in $\alpha=2/L$, $\lambda_M=10$, $\lambda_N=200$,
$\theta_{div}=0.3$, $\theta_{mov}=2$ and $\theta_{del}=0.03$. The solid line
correspond to a nonlinear fitting with $r^2=0.9995$.}
\end{figure*}

\begin{figure*}
\begin{center}
\resizebox{10cm}{4.6cm}{\includegraphics{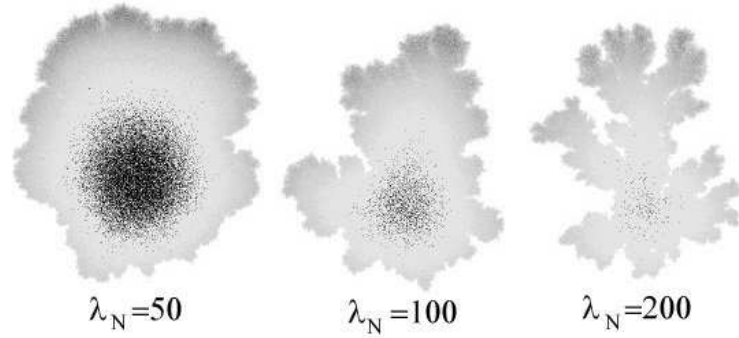}}
\end{center}
\caption{\label{fig:fig4} Simulated growth patterns exhibiting a necrotic core
(in black).  The model parameters are $\alpha=2/L$, $\lambda_M=25$,
$\theta_{div}=0.3$, $\theta_{mov}=\infty$ (without cell migration) and
$\theta_{del}=0.03$. The fraction of necrotic cells is smaller for papillary
patterns than compact ones, suggesting that the optimal growth morphology under
strong nutrients limitation is fractal.}
\end{figure*}

\begin{figure*}
\begin{center}
\resizebox{12.6cm}{6.9cm}{\includegraphics{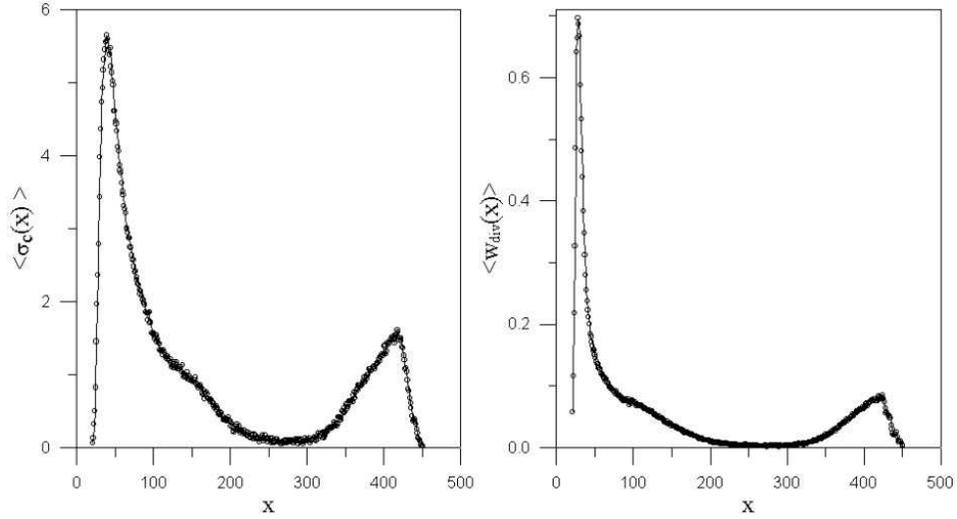}}
\end{center}
\caption{\label{fig:fig5} (a) Density of cancer cells $\sigma_c$ and (b)
division rates $w_{div}$ as a function of the distance from the capillary
vessel along a tumor longitudinal cut. The model parameters were fixed in
$\alpha=2/L$, $\lambda_M=25$, $\lambda_N=50$, $\theta_{div}=0.3$,
$\theta_{mov}=\infty$ (without cell migration) and $\theta_{del}=0.03$. Sharp
maxima at the tumor borders near and opposed to the capillary vessel at $x=0$
are evident. Therefore, the proliferative fraction of cancer cells is
distributed in a thin shell on the tumor border.}
\end{figure*}

\begin{figure*}
\begin{center}
\resizebox{10cm}{10cm}{\includegraphics{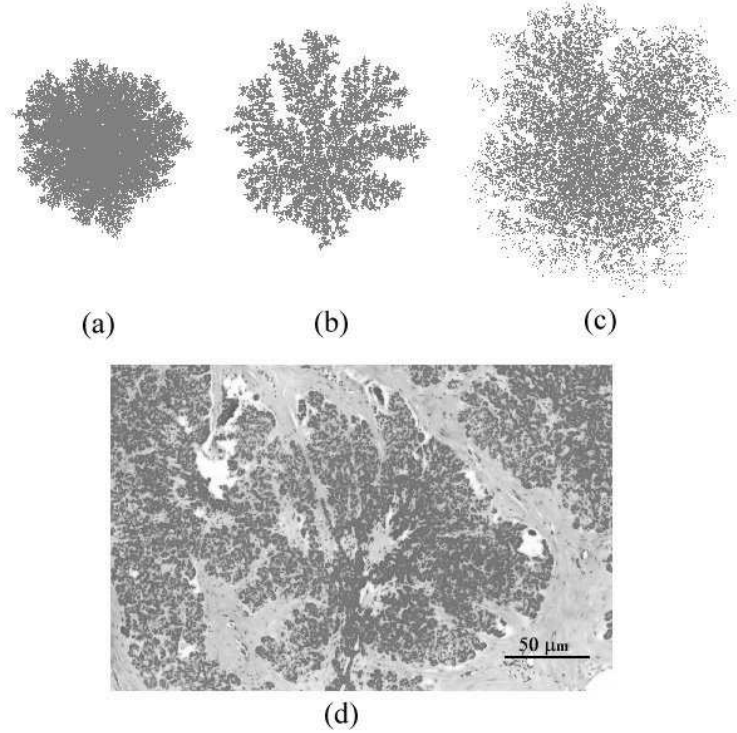}}
\end{center}
\caption{\label{fig:fig6}  Simulated patterns of the nutrient-limited cancer
growth model including the influence among cancer cells mediated by growth
factors.  (a) compact ($k=0.1$ and $\theta_{mov}=1$), (b) disconnected
($k=0.025$ and $\theta_{mov}=0.1$), and (c) ramified ($k=0.025$ and
$\theta_{mov}=1$) morphologies. The remaining parameters of the model were
fixed in $\alpha=3/L$, $\theta_{div}=0.5$, $N^*=\theta_{del}=0.01$, $\lambda=5$
and $\Gamma=10$. The tissue size is $500 \times 500$, with the initial ``cancer
seed'' distant $300$ sites from the capillary and the total number of cancer
cells is $5 \times 10^4$. For comparison, a real ramified pattern observed in
trichoblastoma is shown in (d).}
\end{figure*}

\end{document}